\documentclass[journal]{IEEEtran}
\newcommand{\ee}{\epsilon}			
\usepackage{epsf,psfrag,amssymb,amsfonts,color,cite,fancybox}
\usepackage[mathscr]{eucal}
\usepackage[dvips]{graphicx}
\usepackage{ifpdf}
\usepackage{longtable}	
\usepackage{multicol}
\usepackage{float}

\usepackage{epstopdf}

\ifCLASSINFOpdf

\else

\fi

\usepackage[cmex10]{amsmath}
\usepackage[caption=false,font=footnotesize]{subfig}

\begin{document}
\renewcommand{\thefigure}{\arabic{figure}}
\renewcommand{\thesubfigure}{\alph{subfigure}}

\title{
Long-Term Stability Analysis of Power Systems with Wind Power 
Based on Stochastic Differential Equations: Model Development and Foundations}

\author{Xiaozhe Wang,~\IEEEmembership{Member,~IEEE,}
        Hsiao-Dong Chiang,~\IEEEmembership{Fellow,~IEEE,}
        Jianhui Wang,~\IEEEmembership{Senior Member,~IEEE,}
        Hui Liu,~\IEEEmembership{Member, ~IEEE.}
        Tao Wang,~\IEEEmembership{Member, ~IEEE,}
\thanks{
}
\thanks{Xiaozhe Wang is with the Department of Mechanical Engineering, Massachusetts Institute of Technology, Cambridge,
MA 02138 USA, email: xw264@cornell.edu.}
\thanks{Hsiao-Dong Chiang and Tao Wang are with the School of Electrical and Computer Engineering, Cornell University, Ithaca,
NY 14853 USA, email: hc63@cornell.edu, tw355@cornell.edu}
\thanks{Jianhui Wang is with the Center for Energy, Environmental, and Economic Systems Analysis, Argonne National Laboratory, Argonne, IL 60439 USA, email: jianhui.wang@anl.gov.}
\thanks{Hui Liu is with the School of Electrical and Information Engineering, Jiangsu University, Zhenjiang, China, e-mail: hughlh@126.com.}

}

\maketitle

\begin{abstract}
In this paper, the variable wind power is incorporated into the dynamic model for long-term stability analysis. A theory-based method is proposed for power systems with wind power to conduct long-term stability analysis, which is able to provide accurate stability assessments with fast simulation speed. Particularly, the theoretical foundation for the proposed approximation approach is presented. The accuracy and efficiency of the method are illustrated by several numerical examples.
\end{abstract}


\IEEEpeerreviewmaketitle

\section{Introduction}


Steady increase in load demands and aging transmission networks have pushed many power systems closer to their stability limit \cite{Chiang:book}. Previous practical experience shows that the system may remain stable in the short-term time scale (say 0-30s after the contingency) while becomes unstable in the long-term time scale (say 30s-several minutes after the contingency) due to insufficient reactive power support or poor control schemes \cite{Cutsem:book}-\cite{Andersson:1993}. The focus of long-term stability is on slower and longer duration phenomena after generator dynamics damp out \cite{Kundur:book}-\cite{Kundur:1993}. In this time scale, two-time scale decomposition of the power system dynamic model can be employed \cite{Cutsem:book}\cite{Cutsem:article2000}. Particularly, the dynamics can be classified into short-term dynamics and long-term dynamics based on the time frames of function after contingencies.
As the penetration of wind power continues to grow, the long-term stability of a system may be further affected by the volatile nature of wind power and the distinct dynamic response characteristics of wind turbines \cite{Kling:2001}. Long-term stability analysis deserves more attention to better ensure the secure operation of power grids. It is also crucial to study the impacts of wind power on long-term stability.

In previous literature, impacts of wind power on long-term stability have been discussed in \cite{Vieira:2015}\cite{Bezerra:2012}, which focus on the effects of different control schemes of wind generators. Stability impacts of wind power on transient stability and small signal stability have been discussed in \cite{Rodriguez:2003}-\cite{Loparo:2011}. 
In all above studies, power systems are formulated in ordinary differential equations (ODEs) or differential algebraic equations (DAEs), and wind speed is assumed to be constant without considering the statistic properties of wind. 
To study the effects of the variability of wind power, stochastic differential equations (SDEs) are employed in \cite{crow:2012}-\cite{Nwankpa:2000} for transient stability and small-signal stability. In those studies, the variations induced by wind speed are simply modelled as white noise perturbations on power injections without characterizing the statistic properties of wind speed. In \cite{Chen:2012}, a set-theoretic method is proposed to assess the effect of variability of renewable energies on short-term dynamics under small perturbations.
Regarding long-term stability study of power systems with wind power, the challenges include the stochastic characterization of wind power and the resulting high computational burden in stability assessments, which have not been well addressed. Specifically, introduction of the randomness will clearly increase the computational burden, making the stability assessments of stochastic systems time consuming. 

In this paper, the statistic properties of wind power are well characterized and incorporated in stability analysis; a method is proposed with its theoretical foundation for long-term stability analysis, which is able to reduce the computational burden arising from the randomness. Particularly, a SDE-based model \cite{Milano:2013_1} is utilized to characterize the Weibull-distributed wind speed, which is further incorporated into the complete dynamic model described in SDEs. Under this SDE formulation, a theory-based method, which is to approximate the stochastic model by a deterministic model, is proposed to perform long-term stability analysis. The theoretical foundation for the method is also developed under which the accuracy of the deterministic model is guaranteed. 
Compared to the deterministic models using constant wind speeds, the proposed deterministic model reflects the variable nature of wind power by providing correct stability assessments for the stochastic model. Compared to the stochastic model characterizing the statistic properties of wind speed, the proposed deterministic model takes much less time in time domain simulation.

The rest of the paper is organized as follows. Section \ref{sectionpowermodels} briefly reviews power system models. Section \ref{mathprelim} introduces the preliminaries about the wind speed model by SDEs and singular perturbation method. Section \ref{sectionmodels} presents the formulation of the long-term stability model with wind power based on SDEs. Afterwards, an analytical method with its theoretical foundation is proposed in Section \ref{sectiontheory} for long-term stability analysis of power systems with wind power. Several numerical examples are given in Section \ref{sectionnumerical} to illustrate the feasibility and efficiency of the proposed method. Conclusions and perspectives are stated in Section \ref{sectionconclusion}.



\section{Power System Models}\label{sectionpowermodels}
The deterministic power system long-term stability model, i.e., complete dynamic model, for simulating system dynamic response relative to a disturbance can be described as:
\begin{eqnarray}
\dot{z}_{c}&=&\ee{h}_c({z_c,x,y,z_d})\label{slow ode}\\
\dot{{x}}&=&{f}({z_c,x,y,z_d})\label{fast ode}\\
{0}&=&{g}({z_c,x,y,z_d})\label{algebraic eqn}\\
{z}_d(k)&=&{h}_d({z_c,x,y,z_d(k-1)})\label{slow dde}
\end{eqnarray}

Equation (\ref{slow ode}) describes long-term dynamics including exponential recovery loads, turbine governors (TGs) and over excitation limiters (OXLs), and Eqn (\ref{fast ode}) describes the internal dynamics of devices such as generators, their automatic voltage regulators (AVRs), certain loads such as induction motors, and other dynamically modeled components. Equation (\ref{algebraic eqn}) describes the electrical transmission system and the internal static behaviors of passive devices, and Eqn (\ref{slow dde}) describes long-term discrete events like load tap changers (LTCs) and shunt switchings. 
$h_c$, $f$ and $g$ are continuous functions, and vectors $z_c\in\mathbb{R}^{n_{z_c}}$, ${x}\in\mathbb{R}^{n_x}$ and ${y}\in\mathbb{R}^{n_y}$ are the corresponding long-term/slow state variables, short-term/fast state variables, and algebraic variables. $z_d\in\mathbb{R}^{n_{z_d}}$ are termed as long-term discrete variables whose transitions from $z_d(k-1)$ to $z_d(k)$ depend on system trajectories and occur at distinct times $t_k$ where $k=1,2,3,...N$. 
Besides, $1/\ee$ can be regarded as the maximum time constant among devices.

Load models play an important role in long-term stability analysis, and generally load dynamics appear in Eqn (\ref{algebraic eqn}) as:
\begin{eqnarray}\label{Aleqn5}
p&=&x_p/T_p+p_t\\
q&=&x_q/T_q+q_t\label{Aleqn6}
\end{eqnarray}
where $x_p$ and $x_q$ are internal state variables associated with generic load dynamics. They are described in Eqn (\ref{slow ode}) as:
\begin{eqnarray}\label{Lcdfeqn3}
{\dot{x}_p}&=&-x_p/T_p+p_s-p_t\\
{\dot{x}_q}&=&-x_q/T_q+q_s-q_t\label{Lcdfeqn4}
\end{eqnarray}
where $p_s$ and $p_t$ are the static and transient real power absorptions; similar definition for $q_s$ and $q_t$. $p^0_L$ and $q^0_L$ are PQ load power from power flow solutions\cite{Kundur:book}\cite{Milano:article}.

The tap-changing logic of LTC at time instant $t_k$ shows up in Eqn (\ref{slow dde}), and is given as follows:
\begin{equation}\label{Ldeqn1}
m_{k+1}=\left\{\begin{array}{ll}m_k+\triangle{m}&\mbox{if  }v>v_0+d\mbox{  and  }m_k<m^{max}\\
m_k-\triangle{m}&\mbox{if  }v<v_0-d\mbox{  and  }m_k>m^{min}\\
m_k &\mbox{otherwise}\end{array}\right.
\end{equation}
where $v$ is the controlled voltage of LTC, $v_0$ is the reference voltage, $d$ is half the LTC dead-band, and $m^{max}$ and $m^{min}$ are the upper and lower tap limits \cite{Cutsem:book}\cite{Kundur:book}\cite{Milano:article}.

\section{Mathematical Preliminaries}\label{mathprelim}
In this section, the method to model the Weibull-distributed wind speed by SDE proposed in \cite{Milano:2013_1}\cite{Milano:2013} is briefly reviewed. Singular perturbation theory for ODE \cite{Khalil:book} is also reviewed. 
\subsection{Modelling Wind Speed by SDE}\label{subsection modelling wind by SDE}
Two continuous wind speed models based on SDE have been developed in \cite{Milano:2013_1}\cite{Milano:2013}. The developed models include but not limited to the autocorrelated Weibull distributed wind speed models.
In this paper, model I in \cite{Milano:2013_1} is applied to formulate wind speed. Briefly speaking, memoryless transformation of Ornstein-Uhlenbeck process is used to obtain the Weibull distributed process.

First, consider the following stochastic differential equation with given initial condition $\eta(0)\sim\mathcal{N}(0,\beta^2/2\alpha)$:
\begin{equation}\label{windsde}
\dot{\eta}=-\alpha \eta+\beta\xi
\end{equation}
where $\xi(t)=\frac{dW(t)}{dt}$ is white noise process and $W(t)$ represents a standard Wiener process. The stochastic process $\eta(t)$ is an Ornstein-Uhlenbeck/Gauss-Markov process which is stationary, Gaussian and Markovian with the statistic properties $ \mathrm{E}[\eta(t)]=0$, $\mathrm{Var}[\eta(t)]=\beta^2/2\alpha$, and autocorrelation $\mathrm{Aut}[\eta(t_i),\eta(t_j)]=e^{-\alpha|t_j-t_i|}$.

Then we apply a memoryless transformation to $\eta(t)$ as follows:
\begin{equation}\label{windalg}
\gamma(t)=F_w^{-1}(\Phi(\frac{\eta(t)}{\beta/\sqrt{(2\alpha)}}))
\end{equation}
where $F_w$ is the Weibull cumulative distribution function:
\begin{equation}
F_w(u)=1-e^{(u/\lambda)^k}, \quad \forall u>0
\end{equation}
with the shape parameter $k> 0$ and the scale parameter $\lambda>0$,
and $\Phi$ is the Gaussian cumulative distribution function:
\begin{equation}
\Phi(\frac{u-\mathrm{E}[u]}{\sqrt{\mathrm{Var}[u]}})=\frac{1}{2}(1+\mathrm{erf}(\frac{u-\mathrm{E}[u]}{\sqrt{2\mathrm{Var}[u]}})),\quad \forall u\in\mathbb{R}
\end{equation}
The resulting process $\gamma(t)$ is a Weibull distributed stochastic process with the statistic properties $ \mathrm{E}[\gamma(t)]=\lambda\Gamma(1+\frac{1}{k})=\mu_w$, $\mathrm{Var}[\gamma(t)]=\lambda^2 \Gamma(1+\frac{2}{k})-\mu_w^2$, and $\mathrm{Aut}[\gamma(t_i),\gamma(t_j)]\approx e^{-\alpha|t_j-t_i|}$. Note that $\mathrm{E}[\gamma(t)]$, $\mathrm{Var}[\gamma(t)]$ and $\mathrm{Aut}[\gamma(t_i),\gamma(t_j)]$ do not depend on the parameter $\beta$, i.e., $\beta$ in Eqn (\ref{windsde}) does not affect the statistic properties of stochastic process $\gamma(t)$\cite{Milano:2013_1}. Since wind speed is autocorrelated in the range of hours, $\alpha$ depending on the decay rate of autocorrelation is usually small in the time frame of stability analysis.

Note that although we consider the Weibull-distributed wind speed model in this paper, Eqn (\ref{windsde})-(\ref{windalg}) can be used to formulate other probability distributions by applying different $F_w$ in Eqn (\ref{windalg}). Refer to \cite{Milano:2013_1} for more details.

\subsection{Singular Perturbation Theory for ODE}
We next consider the following general singular perturbed model (slow-fast model):
\begin{eqnarray}\label{singular perturb}
\Sigma_\ee:\dot{z}&=&f(z,x)\qquad z\in\mathbb{R}^{n_z}\\
\ee\dot{x}&=&g(z,x)\qquad x\in\mathbb{R}^{n_x}\nonumber
\end{eqnarray}
where $\ee$ is a small positive parameter. $z$ is a vector of slow variables while $x$ is a vector of fast variables. 

The slow model is obtained by setting $\ee=0$ in (\ref{singular perturb}):
\begin{eqnarray}\label{slow}
\Sigma_0:\dot{z}&=&f(z,x)\qquad z\in\mathbb{R}^{n_{z}}\\
0&=&g(z,x)\qquad x\in\mathbb{R}^{n_x}\nonumber
\end{eqnarray}
The algebraic equation $0=g(z,x)$ constrains the slow dynamics to the \textit{constraint manifold} defined by:
\begin{equation}
\Gamma:=\{(z,x)\in\mathbb{R}^{n_{z}}\times\mathbb{R}^{n_x}:g(z,x)=0\}
\end{equation}
The trajectory of model (\ref{slow}) starting at $(z_0,x_0)$ is denoted by $\phi_0(t,z_0,x_0)$, and the stability region is defined as:
\begin{equation}
A_0(z_s,x_s):=\{(z,x)\in\Gamma: \phi_0(t,z,x)\rightarrow(z_s,x_s)\mbox{ as }t\rightarrow\infty\}\nonumber
\end{equation}

The \textit{singular points} of system (\ref{slow}) or \textit{singularity $S$} is defined as:
\begin{equation}
S:=\{(z,x)\in\Gamma:\mbox{ det}(\partial_xg)(z,x)=0\}
\end{equation}
where $\partial_x g$ is the Jacobian matrix of g with respect to x.

Singular points can drastically influence the trajectories of the DAE system (\ref{slow}). Typically, the singular set $S$ is a set of maximal dimension $(n_z-1)$ embedded in $\Gamma$, and $\Gamma$ is separated by $S$ into open regions \cite{Veukatasubramanian:article}\cite{Alberto:article}.

%

We next define the fast model, i.e., boundary layer model, associated with the singularly perturbed model. Define the fast time scale $\kappa=t/\ee$. In this time scale, model (\ref{singular perturb}) takes the form:
\begin{eqnarray}\label{singular perturb in kappa}
\Pi_\ee:\frac{dz}{d\kappa}&=&{\ee}f(z,x)\qquad z\in\mathbb{R}^{n_z}\\
\frac{dx}{d\kappa}&=&g(z,x)\qquad\quad x\in\mathbb{R}^{n_x}\nonumber
\end{eqnarray}
Let $\ee=0$, we obtain the fast model as follows:
\begin{eqnarray}\label{BLS}
\Pi_f:\frac{dz}{d\kappa}&=&0\qquad \qquad z\in\mathbb{R}^{n_z}\\
\frac{dx}{d\kappa}&=&g(z,x)\qquad x\in\mathbb{R}^{n_x}\nonumber
\end{eqnarray}
where $z$ is frozen and treated as a parameter. The constraint manifold $\Gamma$ is a set of equilibriums of models (\ref{BLS}). For each fixed $z$, a fast dynamical model (\ref{BLS}) is defined.


\noindent\textit{Definition 1: Stable Component of the Constraint Manifold / Uniformly Asymptotically Stable}

Assuming $(z,x)\notin S$, and $x=j(z)$ is an isolated root of equation:
\begin{equation}\label{BLSsep}
0=g(z,x)
\end{equation}
then $x=j(z)$ is an equilibrium point of system (\ref{BLS}) and is also termed as \textit{the slow manifold} of system (\ref{singular perturb}). If $x=j(z)$ is an asymptotically stable equilibrium point of system (\ref{BLS}) for all $z\in Z$, then 
we say that the equilibrium point $j(z)$ is \textit{uniformly asymptotically stable}, and the slow manifold $x=j(z)$ is a \textit{stable component} of the constraint manifold $\Gamma$.

Under the following conditions, the slow model (\ref{slow}) 
provides good approximations of the singular perturbed model (\ref{singular perturb}).
Assume that the equilibrium point $x=j(z)$ of the fast model is uniformly asymptotically stable in $z\in Z$ and the initial condition $(z_0,x_0)$ is in the stability region $A_t(z_0,j(z_0))$ of the initial fast model with $z_0\in Z$, then solution of singular perturbed model (\ref{singular perturb}) will approach $j(z)$ in a time of order $\ee|\mbox{log}\ee|$, and the solution of model (\ref{singular perturb}) remains in an $\ee$-neighbourhood of the slow manifold $j(z)$\cite{Khalil:book}\cite{Gentz:2003}. Furthermore, there exists an invariant manifold $x={j^\star}(z,\ee)=j(z)+O(\ee)$ for sufficiently small $\ee$ 
in $z\in Z$\cite{Khalil:book}\cite{Fenichel:1979}, and the dynamics on invariant manifold is given by:
\begin{equation}\label{slow_ee}
\dot{z}=f({j^\star}(z,\ee),z)
\end{equation}
which can be approximated by regular perturbation theory. Specifically, Eqn (\ref{slow_ee}) is reduced to slow model (\ref{slow}) at $\ee=0$.

\section{sde formulation of power system models}\label{sectionmodels}

When the long-term stability model is represented in $\tau$ time scale, where $\tau=t\ee$, and $\prime$ refers to $\frac{d}{d\tau}$, the long-term stability model of power system can be represented as:
\begin{eqnarray}\label{complete}
{z}_{c}^\prime&=&{h}_c({z_c,x,y,z_d})\label{complete_1}\\
\ee{x}^\prime&=&{f}({z_c,x,y,z_d})\\
{0}&=&{g}({z_c,x,y,z_d})\label{complete_2}\\
z_d(k)&=&h_d(z_c,x,y,z_d(k-1))\label{complete}
\end{eqnarray}

It is possible to decouple the long-term stability model into two decoupled systems (\ref{complete_1})-(\ref{complete_2}) and (\ref{complete}). When discrete variables $z_d$ jump from $z_d(k-1)$ to $z_d(k)$,
$z_d$ are updated first according to (\ref{complete}) and then system (\ref{complete_1})-(\ref{complete_2}) works with fixed parameters $z_d$. In other words, the long-term stability model can be regarded as a collection of continuous DAE systems with fixed discrete parameters. In the remainder of this paper, we focus only on the continuous DAE system (\ref{complete_1})-(\ref{complete_2}). 

When incorporating wind power in the power system models, we use Eqn (\ref{windsde})-(\ref{windalg}) to formulate wind speed. Assuming that there are $n_w$ wind sources in the system, and the wind model is described in $\tau$ time scale, we have
\begin{eqnarray}
\ee\eta_w^\prime&=&-A\eta_w+\sigma\xi=f_w(\eta_w)+\sigma\xi\label{wind}\\
y_w&=&\hat{F}_w^{-1}(\hat{\Phi}(\frac{\eta_w}{\sigma/\sqrt{2A}}))=g_w(\eta_w)\nonumber
\end{eqnarray}
where $\eta_w, y_w \in \mathbb{R}^{n_w}$, $A=\mbox{diag}[\alpha_1,\ldots,\alpha_{n_w}]\in \mathbb{R}^{{n_w}\times{n_w}}$ in which $\alpha_i$ is determined by autocorrelation property of each wind source, $\sigma$ is a small positive parameter, $\int_0^{t}\xi(s)ds$ is a $n_w$-dimensional Wiener process; $\hat{F}_w=[F_w(\eta_{w_1}), F_w(\eta_{w_2}),...F_w(\eta_{w_{n_w}})]^T$, $\hat{\Phi}=[\Phi(\eta_{w_1}),\Phi(\eta_{w_2}),...\Phi(\eta_{w_{n_w}})]^T$, and $g_w:\mathbb{R}^{n_w}\mapsto\mathbb{R}^{n_w}$.

Note that $\sigma$ does not affect statistic properties of $y_w$, which has the same reason that $\beta$ does not affect the statistic properties of $\gamma(t)$ as shown in the last paragraph of Section \ref{subsection modelling wind by SDE}. This property of the formulation will play an important role in the  analysis of Section \ref{sectiontheory}. 


With (\ref{complete_1})-(\ref{complete_2}) and (\ref{wind}), the stochastic long-term stability model can be represented as:
\begin{eqnarray}
{z}_{c}^\prime&=&\bar{h}_c({z_c,x,\bar{y},z_d})\label{sto_cont}\\
\ee{\bar{x}}^\prime&=&\bar{f}({z_c,\bar{x},\bar{y},z_d})+\sigma B {\xi}\nonumber\\
{0}&=&\bar{g}({z_c,\bar{x},\bar{y},z_d})\nonumber
\end{eqnarray}
where $\bar{x}=[x^T,\eta_w^T]^T\in \mathbb{R}^{n_x+n_w}=\mathbb{R}^{n_{\bar{x}}}$, $\bar{y}=[y^T,y_w^T]^T\in \mathbb{R}^{n_y+n_w}=\mathbb{R}^{n_{\bar{y}}}$,
$B=\left[\begin{smallmatrix}0\\I_{n_w}\end{smallmatrix}\right]$.

Under the following assumption, the stochastic DAE system (\ref{sto_cont}) can be simplified as a stochastic ODE system.

\textit{Assumption 1.}
The trajectory of the deterministic long-term stability model (\ref{det_cont}) doesn't meet singularity.
\begin{eqnarray}\label{det_cont}
{z}_{c}^\prime&=&\bar{h}_c({z_c,x,\bar{y},z_d})\\
\ee{\bar{x}}^\prime&=&\bar{f}({z_c,\bar{x},\bar{y},z_d})\nonumber\\
{0}&=&\bar{g}({z_c,\bar{x},\bar{y},z_d})\nonumber
\end{eqnarray}

Under \textit{Assumption 1}, $\bar{y}$ can be expressed by $(z_c,\bar{x},z_d)$ from the algebraic equation as $\bar{y}=l(z_c,\bar{x},z_d)$. Substituting the expression back to model (\ref{sto_cont}), we have:
\begin{eqnarray}\label{sde power dae}
z_c^\prime&=&H_c(z_c,\bar{x},z_d)\\
\ee\bar{x}^\prime&=&F(z_c,\bar{x},z_d)+\sigma B {\xi}\nonumber
\end{eqnarray}
or written in the form:
\begin{eqnarray}\label{sde power}
dz_c&=&H_c(z_c,\bar{x},z_d)d\tau\\
d\bar{x}&=&\frac{1}{\ee}F(z_c,\bar{x},z_d)d\tau+\frac{\sigma}{\sqrt{\ee}}BdW\nonumber
\end{eqnarray}

Besides, system (\ref{det_cont}) can be written as:
\begin{eqnarray}\label{det power}
dz_c&=&H_c(z_c,\bar{x},z_d)d\tau\\
d\bar{x}&=&\frac{1}{\ee}F(z_c,\bar{x},z_d)d\tau\nonumber
\end{eqnarray}

In the next section, we will provide a theoretical foundation for approximating (\ref{sde power}) by (\ref{det power}) in long-term stability analysis. The deterministic model (\ref{det power}) can provide correct stability analysis results for the stochastic model (\ref{sde power}) provided some mild conditions are satisfied. 
Compared to the stochastic model (\ref{sde power}), the deterministic model (\ref{det power}) takes less time in time domain simulation. Compared to other deterministic models using constant wind speed, the deterministic model (\ref{det power}) reflects the variability of the wind speed by providing accurate stability assessments for (\ref{sde power}). 

\section{theoretical basis}\label{sectiontheory}
In this section, 
we apply the singular perturbation theory for SDEs proposed in \cite{Gentz:2006}-\cite{Freidlin:book}, to develop the theoretical foundation for  approximating (\ref{sde power}) by (\ref{det power}). The main results of \cite{Gentz:2006}-\cite{Freidlin:book} are briefly reviewed in Appendix \ref{appendix1}. Note that we have formulated the stochastic long-term stability model (\ref{sde power}) into the conventional form (\ref{sto_singular perburbed}) considered in singular perturbation theory for SDE. Specifically, $Q=0$. We further assume the counter-part of \textit{Assumption 2.1} (see Appendix \ref{appendix1}) in this specific formulation is satisfied. Note that differentiability and uniform boundedness are usually satisfied in real physical systems, and the assumption is reasonable. 

\begin{figure}[!ht]
\centering
\includegraphics[width=2in,keepaspectratio=true,angle=0]{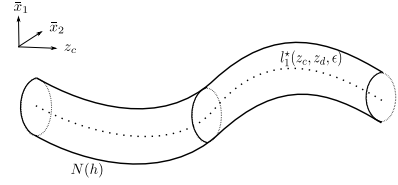}\caption{Illustration of $N(h)$ in which case $n_z=1$, $n_{\bar{x}}$=2. $N(h)$ is a ellipsoidal layer surrounding the invariant manifold $l_1^\star(z_c,z_d,\ee)$. }\label{M(h)}
\end{figure}

We denote the slow manifold of system (\ref{det power}) as $\bar{x}=l_1(z_c,z_d)$. If $\bar{x}=l_1(z_c,z_d)$ is a stable component of the constraint manifold when $z_c\in D_{z_c}\subset\mathbb{R}^{z_c}$, 
then there exists an invariant manifold of system (\ref{det power}) that $\bar{x}=l_1^\star(z_c,z_d,\ee)=l_1(z_c,z_d)+O(\ee)$. We further define the ellipsoidal layer $N(h)$ surrounding the invariant manifold as below:
\begin{eqnarray}
N(h)&=&\{(z_c,\bar{x},z_d):\langle (\bar{x}-l_1^\star(z_c,z_d,\ee)),\nonumber\\
&&L_1^\star(z_c,z_d,\ee)^{-1}(\bar{x}-l_1^\star(z_c,z_d,\ee))\rangle< h^2 \}
\end{eqnarray}
where $\langle\rangle$ denotes the operator of inner product. Note that the matrix $L_1^\star(z_c,z_d,\ee)$ describing the cross section of $N(h)$ is well defined as shown in Appendix \ref{appendix2}. In addition, an illustration of $N(h)$ is shown in Fig. \ref{M(h)}. 

\textit{Theorem 1} shows that under some mild conditions,
the sample paths (i.e., trajectories) of the stochastic long-term stability model (\ref{sde power}) are concentrated in the ellipsoidal layer $N(h)$ surrounding the invariant manifold $l_1^\star(z_c,z_d,\ee)$ of the deterministic long-term stability model (\ref{det power}).







\textit{Theorem 1:} Consider the systems (\ref{sde power})-(\ref{det power}), for some fixed $\ee_0>0$, $h_0>0$, there exist $\delta_0>0$, a time $\tau_1$ of order $\ee|\mbox{log}h|$ such that if the following two conditions are satisfied:

\begin{description}
\item [\textit{i.}] The slow manifold $\bar{x}=l_1(z_c,z_d)$ where $z_c\in D_{z_c}\subset\mathbb{R}^{n_{z_c}}$ is a stable component of the constraint manifold;
\item [\textit{ii.}] The initial condition $(z_c(0),\bar{x}(0),z_d)$ of the stochastic long-term stability model (\ref{sde power}) where $z_c(0)\in D_{z_c}$ satisfies $(z_c(0),\bar{x}(0),z_d)\in N(\delta_0)$, 
\end{description}

then for $\tau\geq\tau_1$, the sample path $(z_c(\tau),\bar{x}(\tau),z_d)$ of the stochastic long-term stability model (\ref{sde power}) satisfies  the following relation
\begin{eqnarray}\label{theorem2}
&&\mathbb{P}\{\exists s\in[\tau_1,\tau]:(z_c(\tau),\bar{x}(\tau),z_d)\notin N(h)\}\nonumber\\
&&\leq C_{n_{z_c},n_x}(\tau,\ee)e^{\frac{-h^2}{2\sigma^2}(1-O(h)-O(\ee))}
\end{eqnarray}
for all $\ee\leq\ee_0$, $h\leq h_0$, where the coefficient $C_{n_{z_c},n_x}(\tau,\ee)=[C^{n_{z_c}}+h^{-n_x}](1+\frac{\tau}{\ee^2})$ is linear time-dependant.

Proof: It can be directly deducted from \textit{Theorem 3} stated in Appendix \ref{appendix1}, hence is omitted.

Note that the detailed expression of $C_{n_{z_c},n_x}(\tau,\ee)$ is omitted here for brevity, which can be obtained directly from Theorem 2.4 in \cite{Gentz:2003}. One key point is that the coefficient is independent of $h$ and $\sigma$, so the probability of the sample path to leave the layer $N(h)$ decays exponentially as $h$ increases. 


One important implication of the theorem is that as soon as we take $h$ slightly larger than $\sigma$, say $h=O(\sigma|\mbox{log}\sigma|)$, the right hand side of Eqn (\ref{theorem2}) becomes very small unless we wait for long time spans. That means the sample paths of the stochastic long-term stability model (\ref{sde power}) are concentrated in $N(\sigma)$. Furthermore, if $h\gg \sigma$ in Eqn (\ref{theorem2}), sample pathes of the stochastic model unlikely leave $N(h)$ as long as slow dynamics permit. Hence, we can explore the trajectory relations between the stochastic model (\ref{sde power}) and the deterministic model (\ref{det power}) without concerning probabilities.  
On the other hand, the condition that $h\gg \sigma$ can be easily satisfied in this SDE formulation of the power system model. As stated before, $\sigma$ doesn't affect the statistic properties of wind speed, thus we can choose $\sigma$ as small as desired such that any satisfactory depth $h$ of the layer satisfies $h\gg\sigma$.

Next we study the trajectory relations between the stochastic long-term stability model (\ref{sde power}) and deterministic long-term stability model (\ref{det power}) under the condition that $h\gg \sigma$, i.e., the depth $h$ of layer $N(h)$ is much larger than the small positive parameter $\sigma$ associated with wind speeds. 
According to singular perturbation theorems\cite{Khalil:book}\cite{Lobry:article2} and the theoretical foundation for the Quasi Steady-State (QSS) model proposed in \cite{Wangxz:CAS}, we have that if the slow manifold $l_1(z_c,z_d)$ of the deterministic long-term stability model is a stable component of the constraint manifold, and the initial point on trajectory of the deterministic long-term stability model lies inside the stability region of the initial short-term stability model, then the trajectory of the long-term stability model will approach the slow manifold in a time of order $\tau_1=\ee|\mbox{log}\ee|$, and moves along the invariant manifold $l_1^\star(z_c,z_d,\ee)$. Refer to \cite{Wangxz:CAS} for more details. 

%
We denote the solution of stochastic long-term stability model (\ref{sde power}) as $(z_{c}(\tau),\bar{x}(\tau),z_d)$, and denote the solution of deterministic long-term stability model (\ref{det power}) as $(z_{cD}(\tau),\bar{x}_D(\tau),z_d)$. Then the following theorem describes the trajectory relations between the two models.

\textit{Theorem 2 (Trajectory Relationship):} Assuming that $h\gg\sigma$, consider the systems (\ref{sde power})-(\ref{det power}), for some fixed $\ee_0\in(0,h)$, there exist $\delta_0>0$, a time $\tau_1$ of order $\ee|\mbox{log}h|$, and $\tau_2>\tau_1$ such that if the following conditions are satisfied:

\begin{description}
\item [\textit{i.}] The slow manifold $\bar{x}=l_1(z_c,z_d)$ where $z_c\in D_{z_c}\subset\mathbb{R}^{n_{z_c}}$ is a stable component of the constraint manifold;
\item [\textit{ii.}] The initial condition of stochastic system (\ref{sde power}) $(z_c(0),\bar{x}(0),z_d)$ where $z_c(0)\in D_{z_c}$ satisfies $(z_c(0),\bar{x}(0),z_d)\in N(\delta_0)$; 
\item [\textit{iii.}] The initial condition $(z_{c}(0), \bar{x}_D(0),z_d)$ of the deterministic long-term stability model (\ref{det power}) is inside the stability region of the initial short-term stability model,
\end{description}

then for $\tau\in[\tau_1,\tau_2]$, the following relations hold: 
\begin{eqnarray}
|\bar{x}(\tau)-\bar{x}_D(\tau)|&=&O(\sigma)\label{corollary3_1}\\
|z_{c}(\tau)-z_{cD}(\tau)|&=&O(\sigma\sqrt{\ee})\label{corollary3_2}
\end{eqnarray}
for all $\ee\in(0,\ee_0)$.

%

Proof: See Appendix \ref{appendix3}.

\begin{figure}[!ht]
\centering
\includegraphics[width=3in,keepaspectratio=true,angle=0]{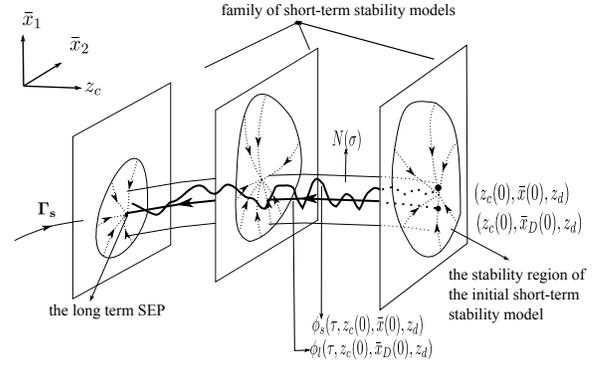}\caption{The trajectory $\phi_s(\tau,z_c(0),\bar{x}(0),z_d)$ of the stochastic long-term stability model is concentrated in $N(\sigma)$, and can be approximated by trajectory $\phi_l(\tau,z_c(0),\bar{x}_D(0),z_d)$ of the deterministic long-term stability model. }\label{stochasticdeterministic}
\end{figure}

\textit{Theorem 2} indicates that trajectory of the stochastic long-term stability model is concentrated in $N(\sigma)$, and can be approximated by trajectory of the deterministic long-term stability model as shown in Fig. \ref{stochasticdeterministic}, provided that the required conditions are satisfied. On the other hand, these conditions are mild and are usually satisfied in practical applications, hence, the deterministic long-term stability model can be applied for stability analysis in most situations. 

\section{Numerical Illustration}\label{sectionnumerical}
In this section, three examples are given to show that the deterministic long-term stability model can provide accurate approximations and correct stability assessment for the stochastic long-term stability model, provided that all conditions of \textit{Theorem 2} are satisfied. At the same time, the time consumed by the deterministic long-term stability model is much less than that consumed by the stochastic long-term stability model. All simulations were performed on PSAT-2.1.8\cite{Milano:article}. Note that both the stochastic model and the deterministic model applied fixed time step size, whereas the efficiency of the deterministic model can be further improved by using larger or adaptive step size.

\subsection{Numerical Example I}
The first example is a 14-bus system in which there are two wind sources and two doubly-fed induction generators (DFIGs) at Bus 2 and Bus 8 respectively. The wind speeds are characterized by Weibull distributed stochastic processes with parameters $k=1.51$, $\lambda=3.36$ for wind source at Bus 1 and $k=5$, $\lambda=20$ for the other one. Both of them have autocorrelation parameter $\alpha=0.2575/3600$\cite{Milano:2013_1}. In addition, there are three synchronous generators (SYNs) which are controlled by AVRs, TGs and OXLs. And there is an exponential recovery load at Bus 5 and one continuous LTC between Bus 4-9 whose initial time delay is 20s. At 1s, there is one line loss between Bus 10-9. 

\begin{figure}[!ht]
\centering
\begin{minipage}[t]{0.5\linewidth}
\includegraphics[width=1.6in ,keepaspectratio=true,angle=0]{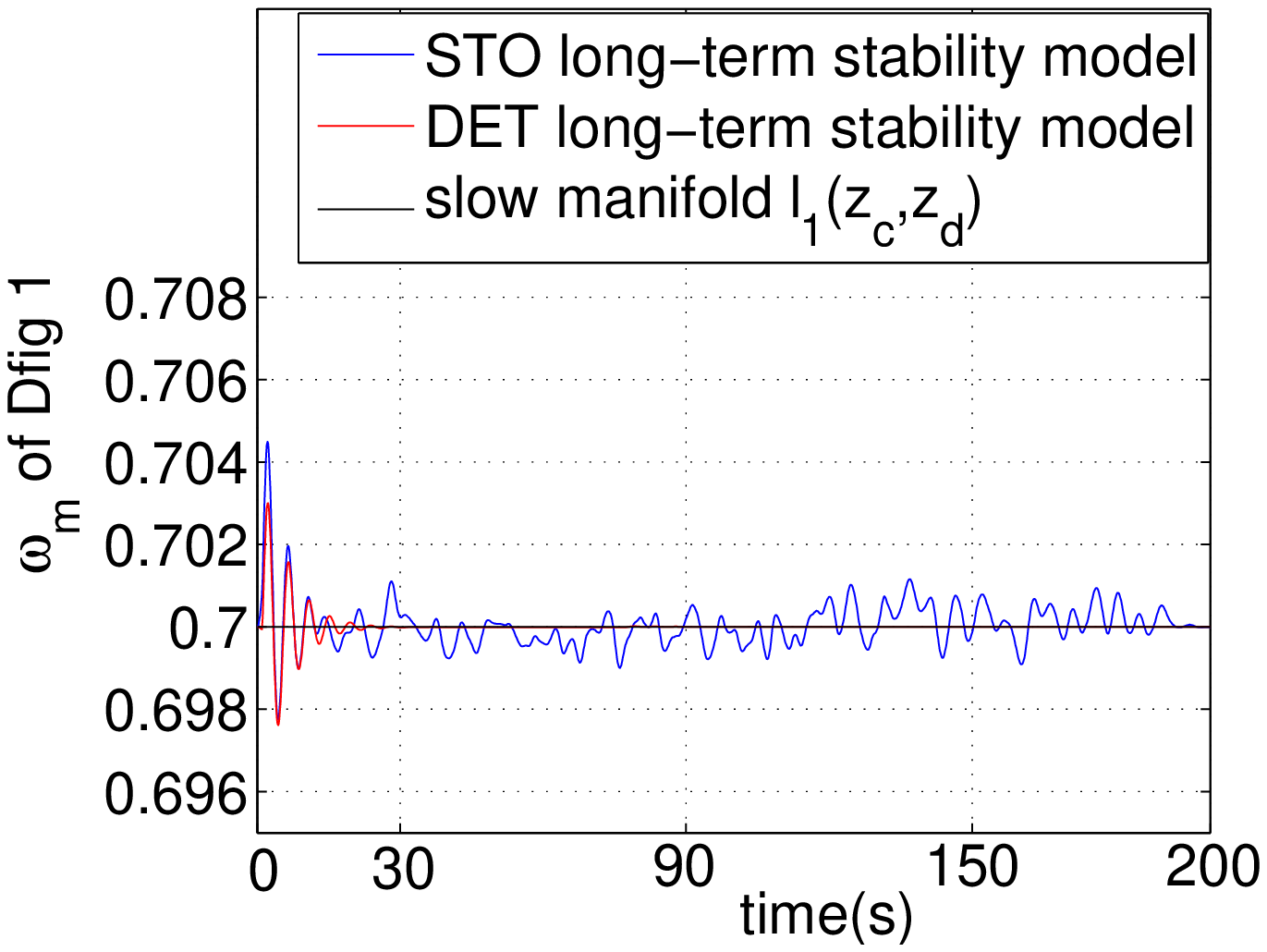}
\end{minipage}%
\begin{minipage}[t]{0.5\linewidth}
\includegraphics[width=1.6in ,keepaspectratio=true,angle=0]{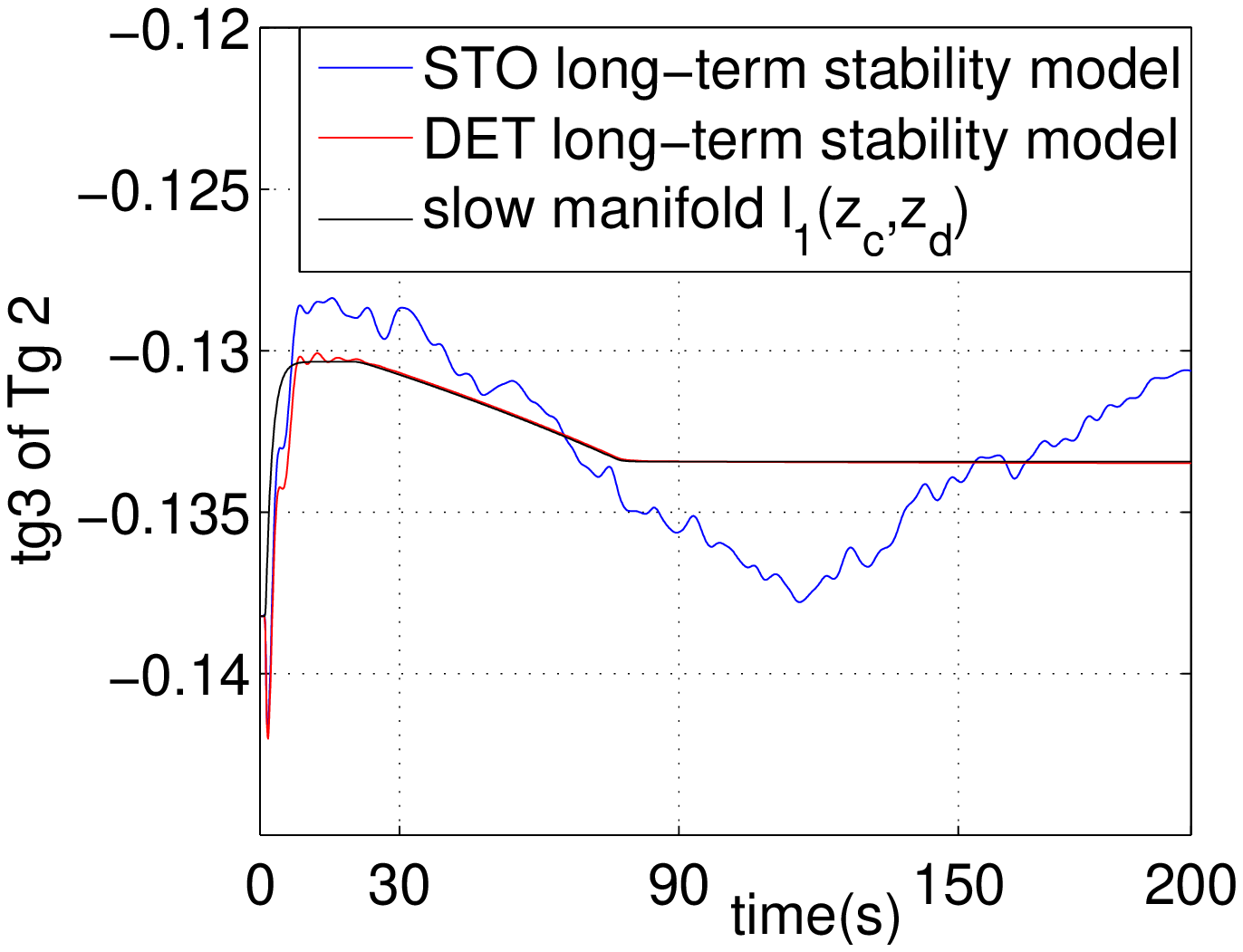}
\end{minipage}
\begin{minipage}[t]{0.5\linewidth}
\includegraphics[width=1.6in ,keepaspectratio=true,angle=0]{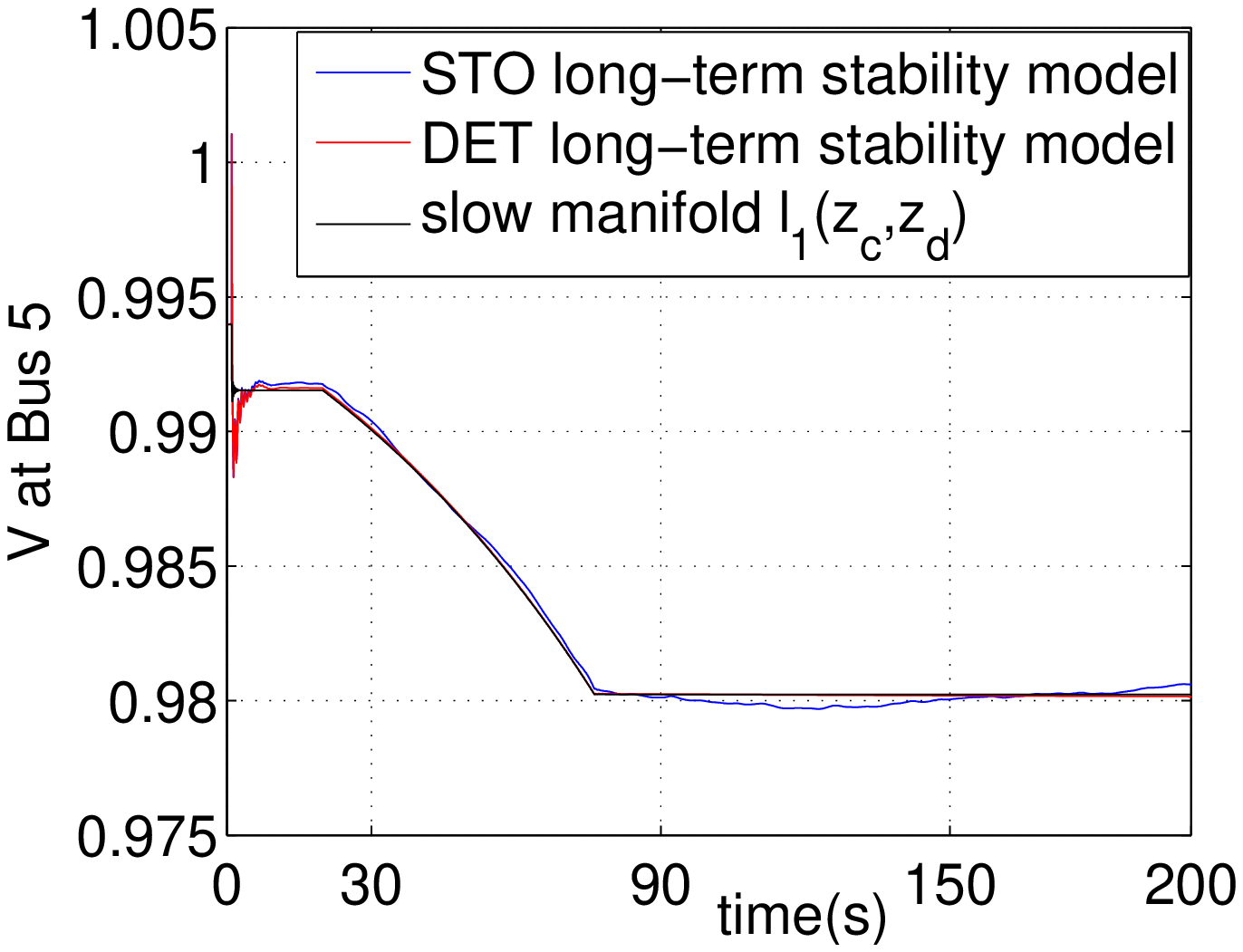}
\end{minipage}%
\begin{minipage}[t]{0.5\linewidth}
\includegraphics[width=1.6in ,keepaspectratio=true,angle=0]{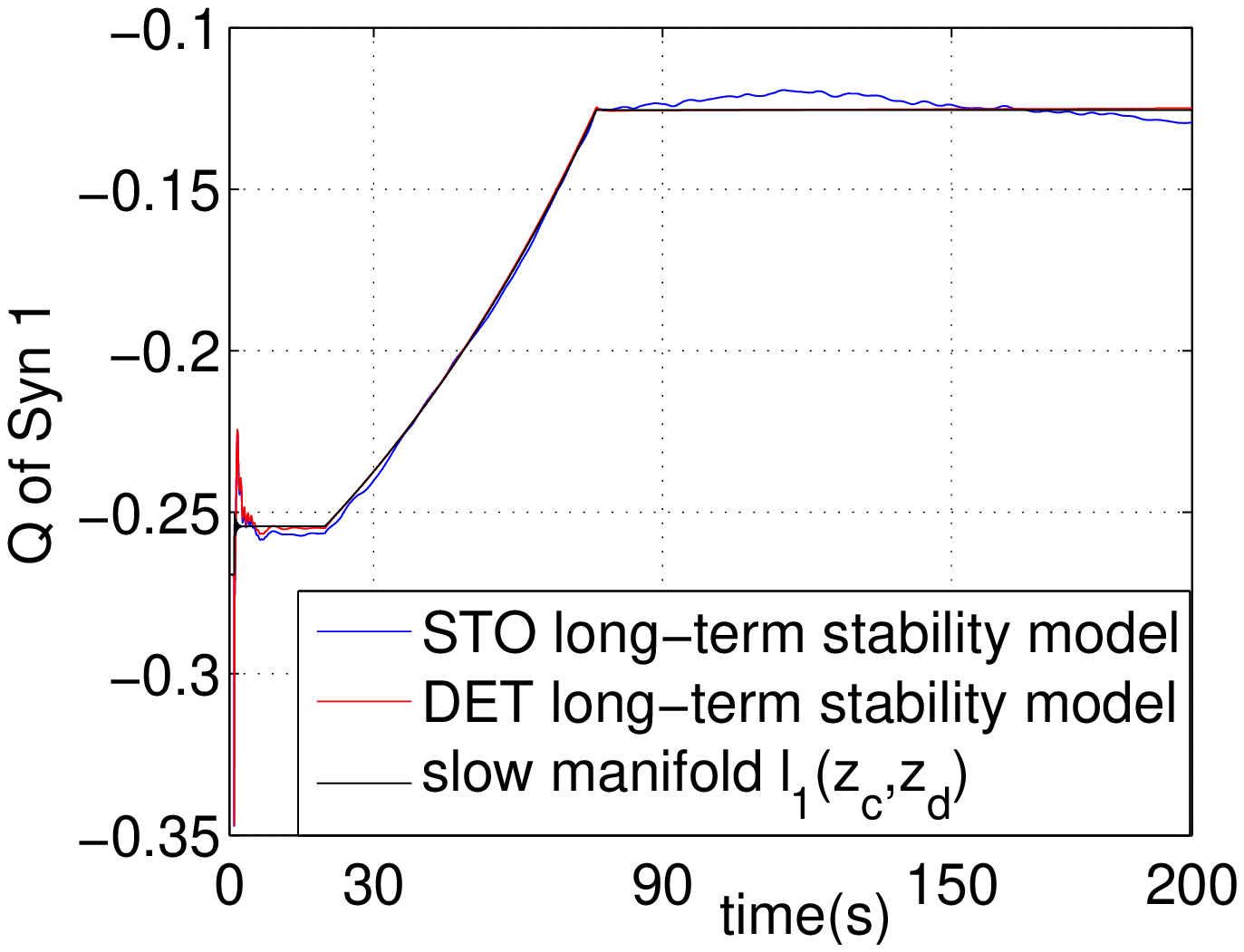}
\end{minipage}
\caption{The trajectory comparisons of the stochastic long-term stability model and deterministic long-term stability model. Since all conditions of \textit{Theorem 2} are satisfied, the deterministic long-term stability model provides accurate approximations in terms of trajectories and stability assessment.}\label{my14wind1}
\end{figure}

One of the sample paths of the stochastic model took 156.96s to simulate, while the deterministic model only took 93.78s. The time required by one sample path has already been longer than that of the deterministic model, let alone the cases where several sample paths of the stochastic model are required to evaluate the stability of the system. The comparison between trajectories of the stochastic long-term stability model and those of the deterministic long-term stability model is shown in Fig. \ref{my14wind1}. Slow manifold of the deterministic model which is obtained from the Quasi Steady-State model \cite{Cutsem:book} is also denoted. Trajectory of the stochastic long-term stability model is fluctuating around that of the deterministic long-term stability model due to variable wind speed, while it always keeps a small distance to trajectory the deterministic long-term stability model. The stability assessments of both models are the same that the system is long-term stable. In this case, the deterministic model doesn't meet singularity as seen from Fig. \ref{my14wind1} and the slow manifold of the deterministic model is a stable component of the constraint manifold; trajectory of the deterministic long-term stability model moves along the invariant manifold $l_1^\star(z_c,z_d,\ee)$ which is a $\ee$-neighbourhood of the slow manifold, thus we can apply the results of \textit{Theorem 2}.  As a result, the deterministic long-term stability model provides accurate approximations for the stochastic long-term stability model with correct stability assessment that the system is long-term stable.


\subsection{Numerical Example II}
The second example is also a 14-bus system. There is one wind source and a corresponding DFIG at Bus 2. The Weibull distributed wind speed is characterized by parameters $k=1.51$, $\lambda=3.36$ and $\alpha=0.2575/3600$. In addition, there are four generators which are controlled by AVRs, TGs and OXLs. And there are three exponential recovery loads at Bus 9, Bus 10 and Bus 14 respectively. There are two continuous LTCs between Bus 4-9 and Bus 14-13, whose initial time delays are 30s. At 1s, there are three line losses between Bus 6-13, Bus 7-9 and Bus 6-11. 

\begin{figure}[!ht]
\centering
\begin{minipage}[t]{0.5\linewidth}
\includegraphics[width=1.6in ,keepaspectratio=true,angle=0]{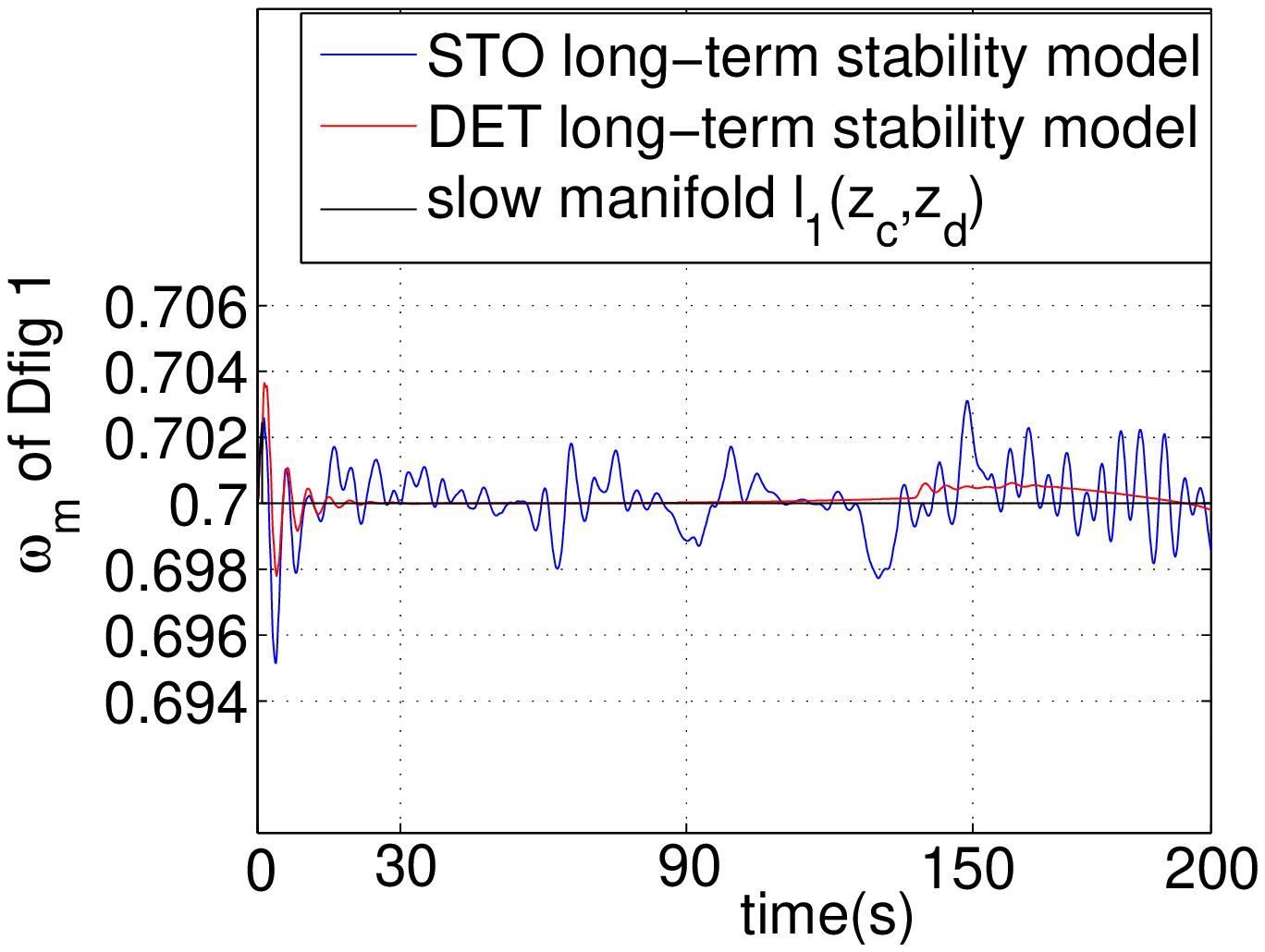}
\end{minipage}%
\begin{minipage}[t]{0.5\linewidth}
\includegraphics[width=1.6in ,keepaspectratio=true,angle=0]{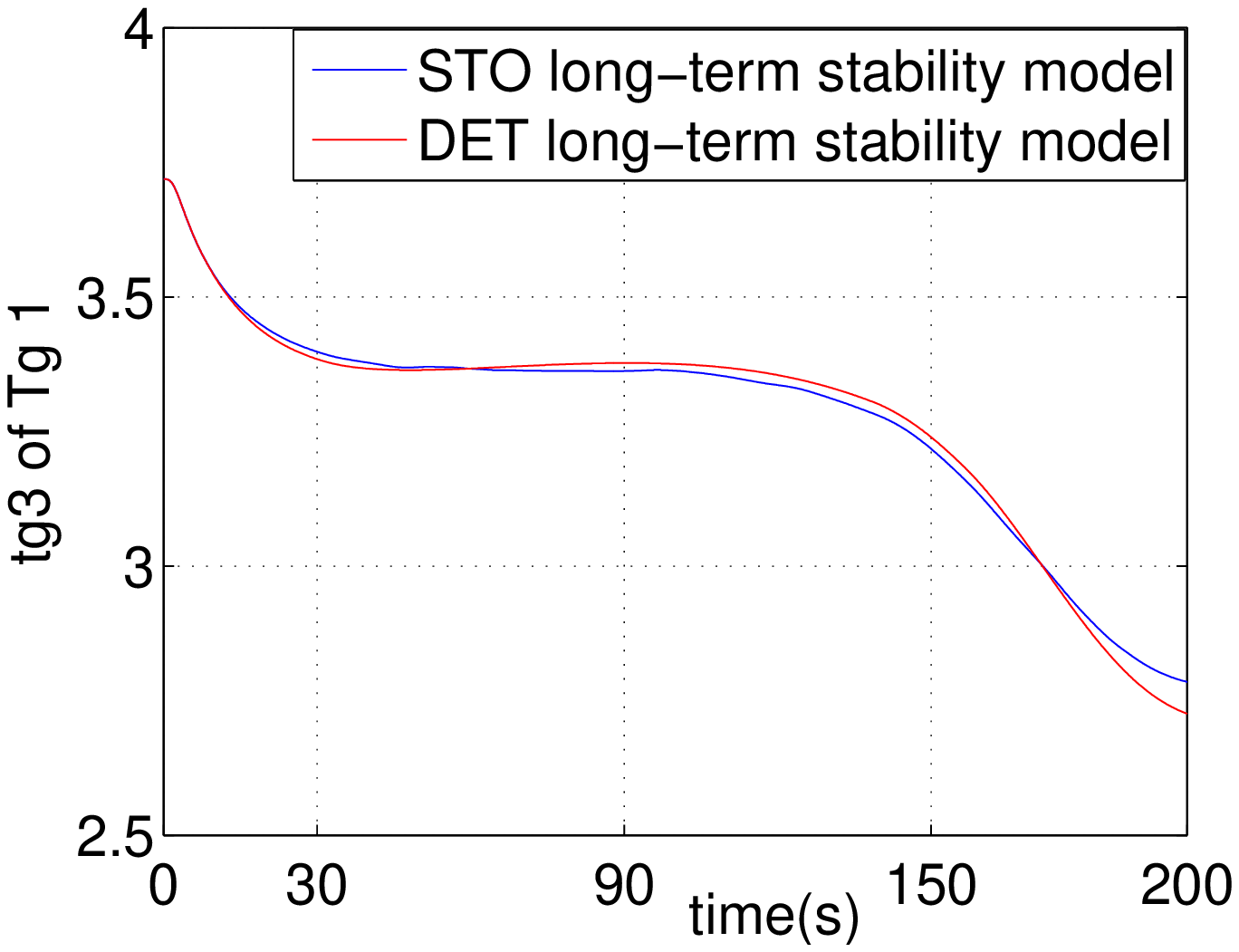}
\end{minipage}
\begin{minipage}[t]{0.5\linewidth}
\includegraphics[width=1.6in ,keepaspectratio=true,angle=0]{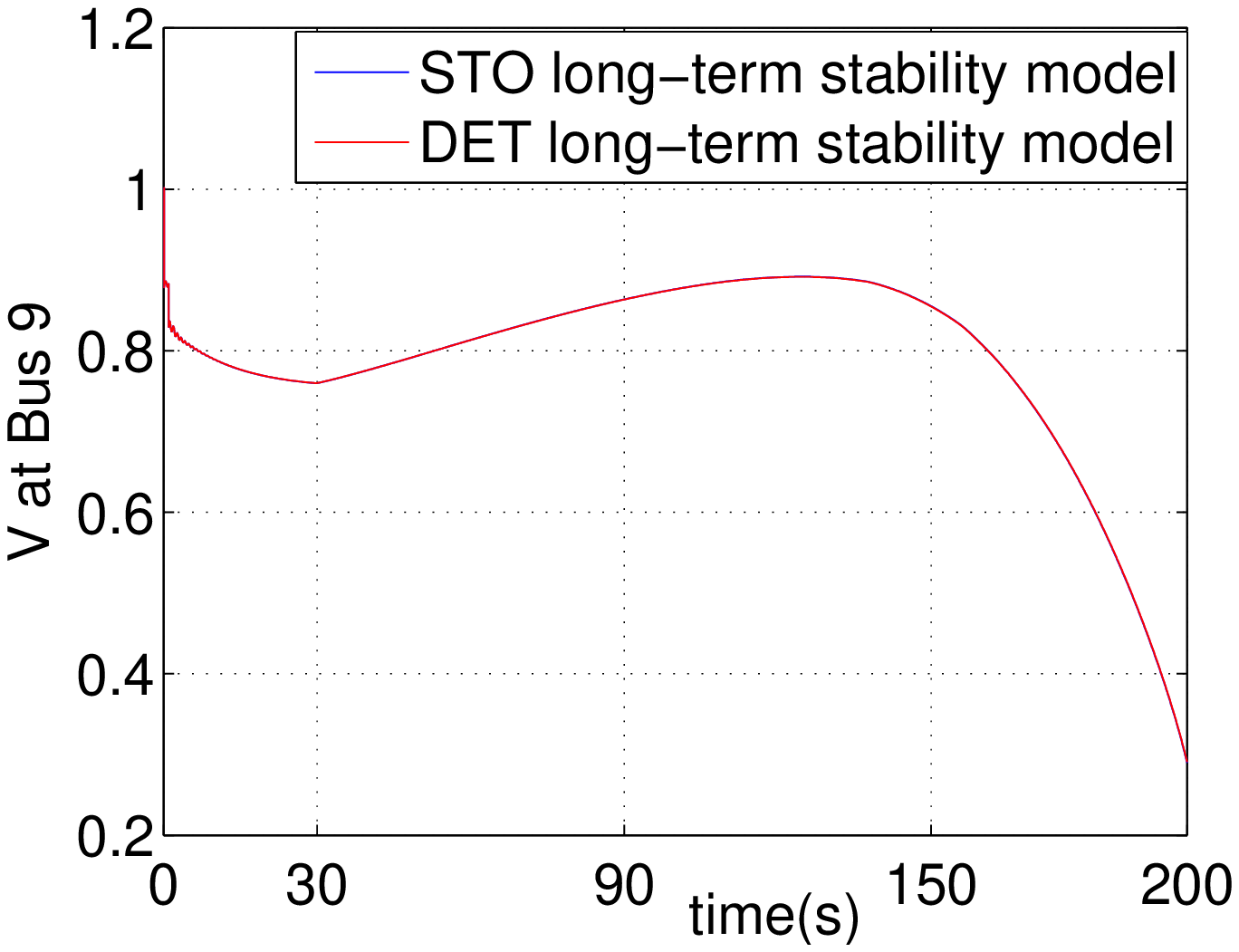}
\end{minipage}%
\begin{minipage}[t]{0.5\linewidth}
\includegraphics[width=1.6in ,keepaspectratio=true,angle=0]{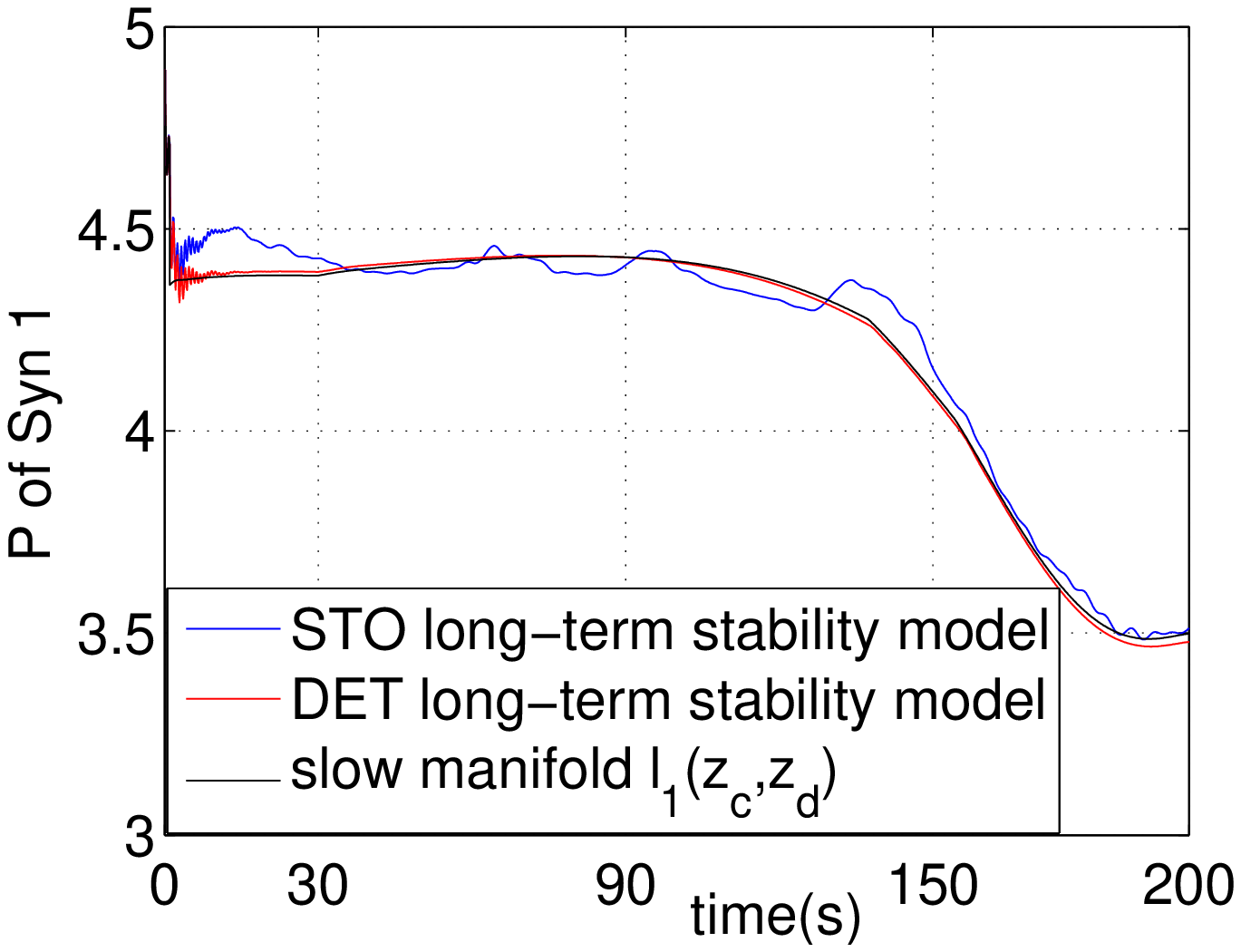}
\end{minipage}
\caption{The trajectory comparisons of the stochastic long-term stability model and the deterministic long-term stability model. Since all conditions of \textit{Theorem 2} are satisfied, the deterministic long-term stability model provides accurate approximations in terms of trajectories and stability assessment that the system is long-term unstable.}\label{my14trywind3}
\end{figure}

In this case, both the stochastic long-term stability model and the deterministic long-term stability model have voltage collapse, and the deterministic long-term stability model successfully captures the voltage instability. One of the sample paths of the stochastic model took 120.86s to simulate, while the deterministic model took 90.64s. The comparison between the trajectories of the stochastic long-term stability model and those of the deterministic long-term stability model as well as the slow manifold are shown in Fig. \ref{my14trywind3}. Similarly, as all conditions of \textit{Theorem 2} are satisfied, the deterministic long-term stability model provides accurate approximations for the stochastic long-term stability model with correct stability assessment.


Physically speaking, as LTCs try to recover voltages at Bus 9 and Bus 13, OXLs at Generator 2 and Generator 4 are activated. LTCs require more power support from generators, however, OXLs protect generators from overheating, thus restrict power output. As the system can not meet the reactive power demand imposed by LTCs, voltages finally collapse.

\subsection{Numerical Example III}
The last example is a larger system which has 145 buses\cite{testcasearchive}.  There are 6 DFIGs driven by different wind sources. Besides, there are 50 synchronous generators each of which is controlled by AVR, and Generator 10-Generator 20 are also controlled by TGs. There are OXLs at Generator 1-Generator 6 whose initial time delays are 50s. Additionally, there are 5 continuous LTCs between Bus 1-33, Bus 1-5, Bus 79-95, Bus 79-92, Bus 60-95 respectively. All LTCs have initial time delays of 50s. At 0.5s, there are three line losses including Bus 94-95, Bus 94-138 and Bus 95-138. 

\begin{figure}[!ht]
\centering
\begin{minipage}[t]{0.5\linewidth}
\includegraphics[width=1.6in ,keepaspectratio=true,angle=0]{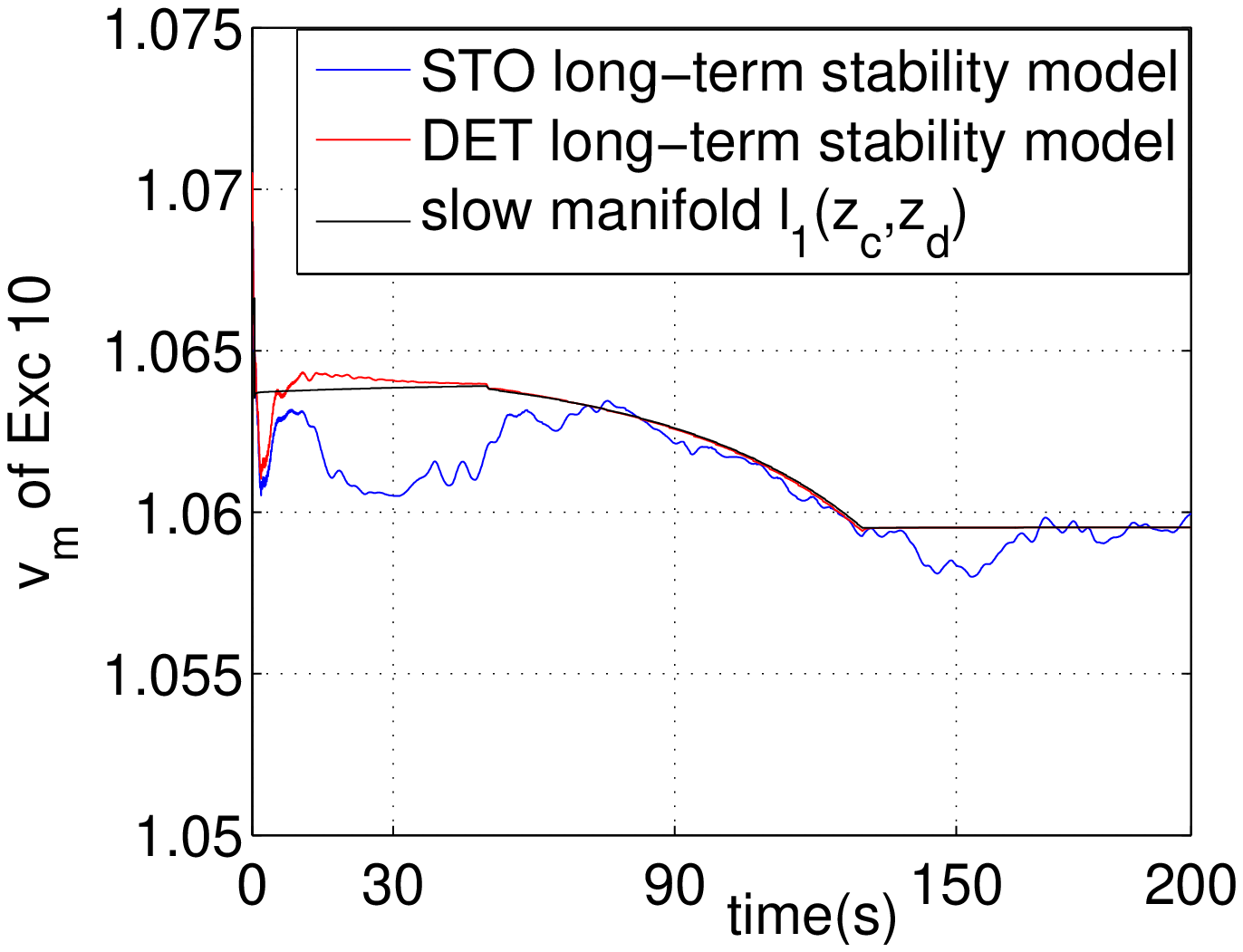}
\end{minipage}%
\begin{minipage}[t]{0.5\linewidth}
\includegraphics[width=1.6in ,keepaspectratio=true,angle=0]{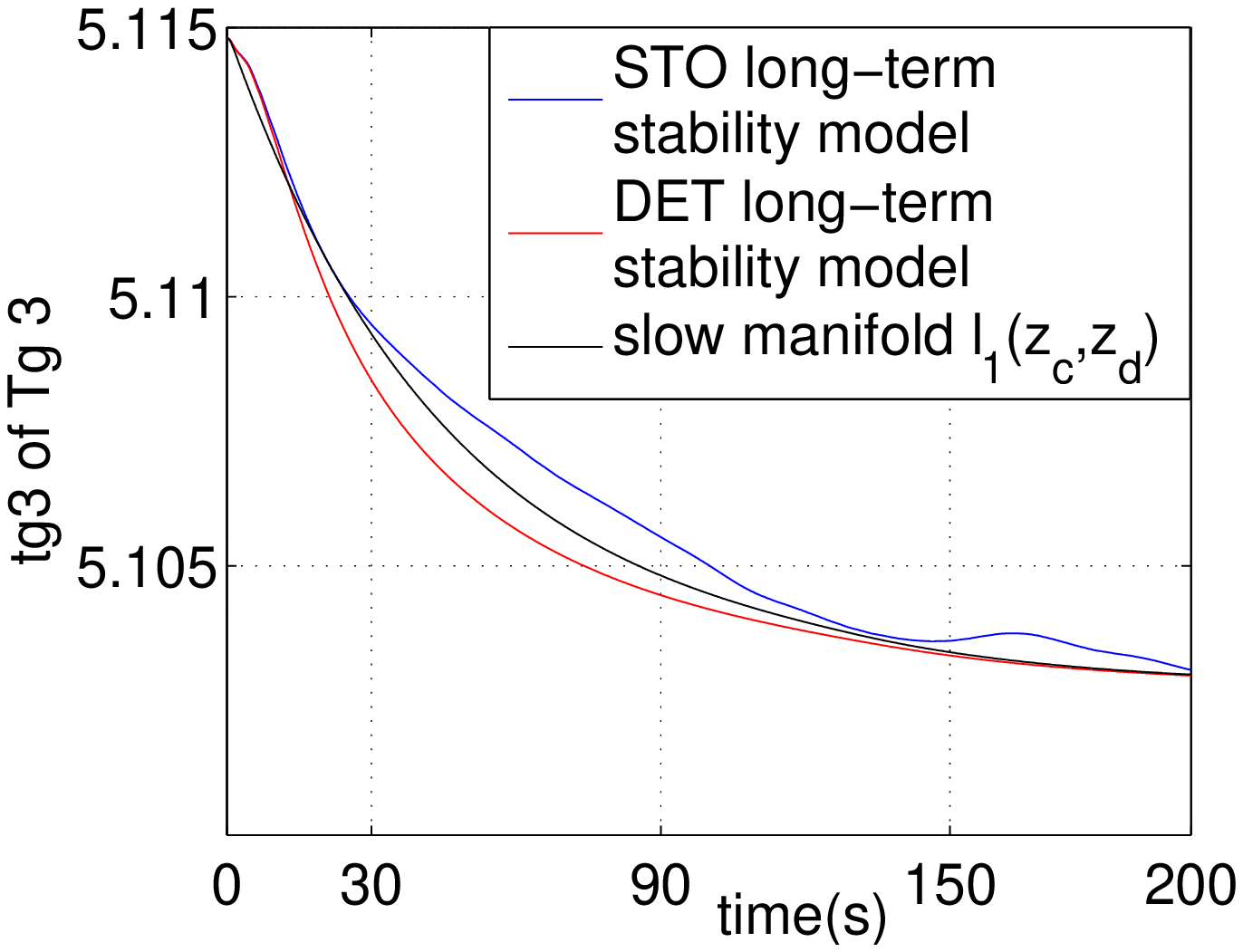}
\end{minipage}
\begin{minipage}[t]{0.5\linewidth}
\includegraphics[width=1.6in ,keepaspectratio=true,angle=0]{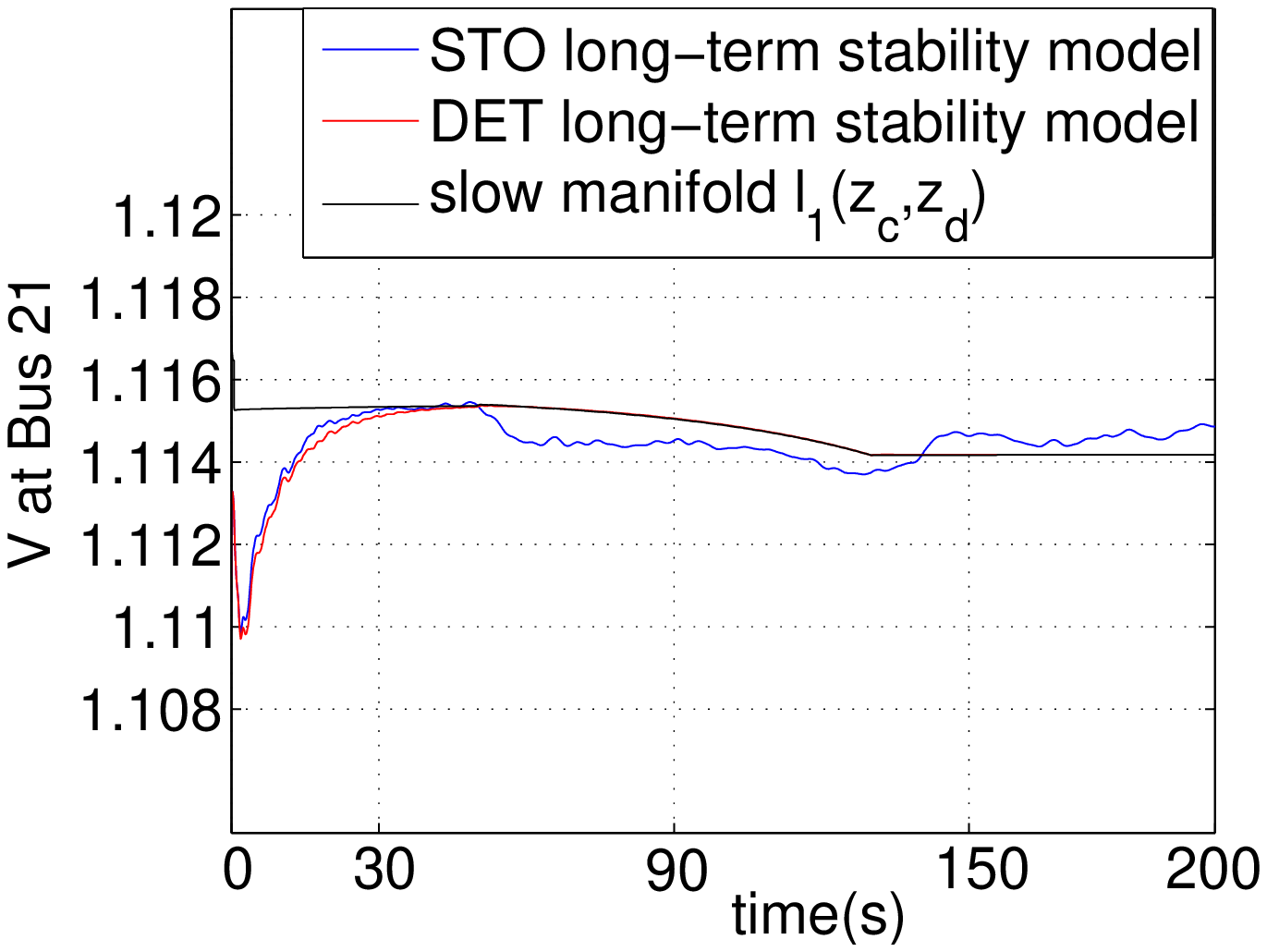}
\end{minipage}%
\begin{minipage}[t]{0.5\linewidth}
\includegraphics[width=1.6in ,keepaspectratio=true,angle=0]{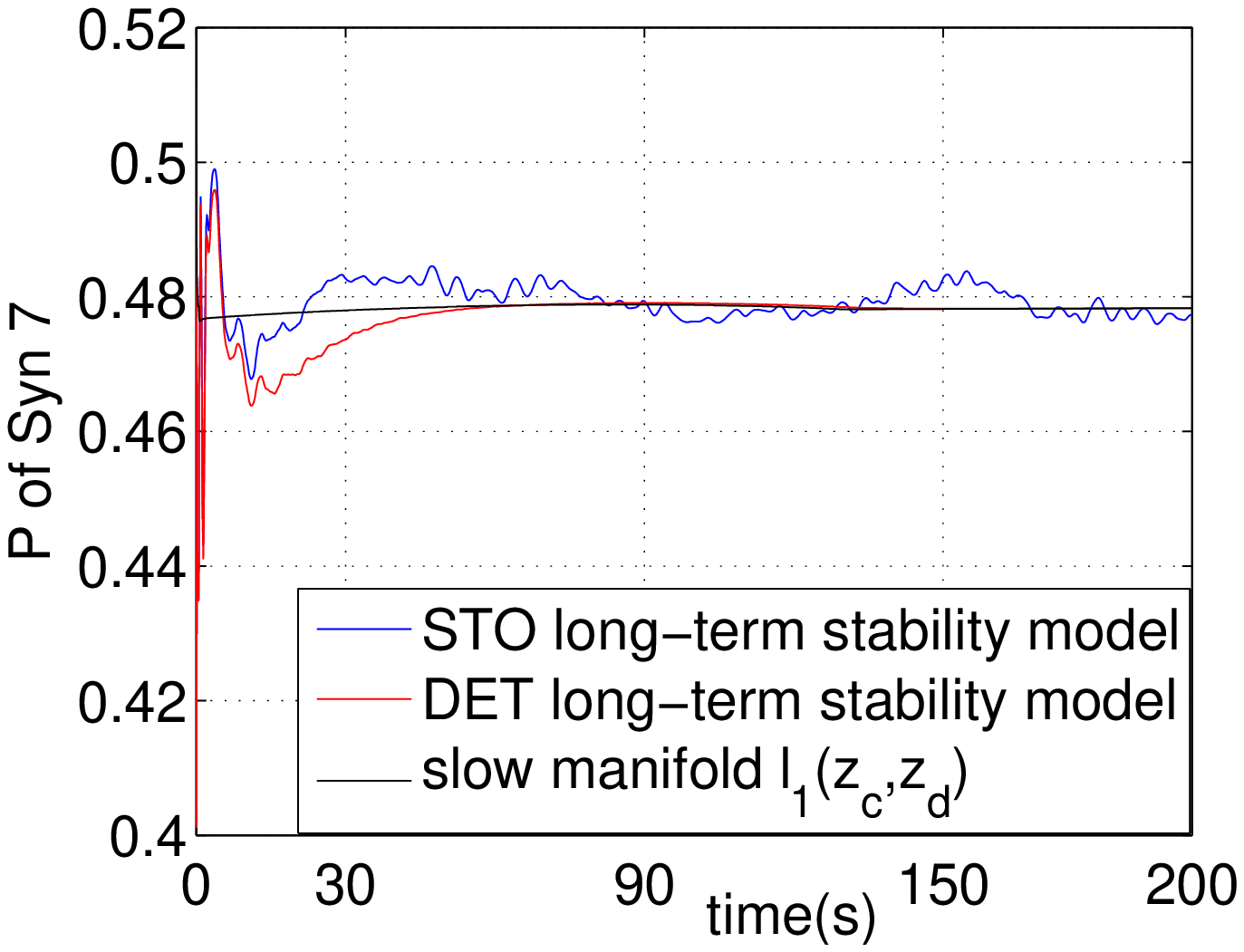}
\end{minipage}
\caption{The trajectory comparisons of the stochastic long-term stability model and the deterministic long-term stability model. The deterministic long-term stability model provides accurate approximations for both trajectories and stability assessment.}\label{my145wind}
\end{figure}


The stochastic long-term stability model took 145.71s to simulate, while the deterministic long-term stability model took 91.18s. The comparison of the trajectories of the stochastic long-term stability model and those of the deterministic long-term stability model is shown in Fig. \ref{my145wind}. Slow manifold of the deterministic model is also labelled. 
The simulation results show that 
the deterministic long-term stability model is able to provide accurate approximations for the stochastic long-term stability model with correct stability assessment, as all conditions of \textit{Theorem 2} are satisfied.

\section{Conclusions and Perspectives}\label{sectionconclusion}
In this paper, we have incorporated the Weibull-distributed model of wind speed proposed in \cite{Milano:2013_1} into the power system dynamic model for long-term stability analysis. Under the SDE formulation, we have theoretically shown that the stochastic model can be approximated by a corresponding deterministic model for stability analysis under mild conditions. The proposed deterministic model takes much less time than the stochastic model, while is able to provide accurate stability assessments for the stochastic model. 

This paper may represent the first attempt to apply the singular perturbation method for stochastic differential equations in power system stability research. The proposed methodology seems promising for stability analysis as the analytical results developed are not limited by the statistic properties of wind power, specific system models, penetration level and so forth. More particularly, the proposed method can accommodate wind speed models with different distributions like Rayleigh distribution, Lognormal distribution, Gamma distribution, etc. 
As a systematic method to model power systems as SDEs has been proposed in \cite{Milano:2013}, we believe that the analytical results in this paper can be readily generalized, and thus provide a unified framework to conduct stability analysis for systems with uncertainties arising from load variations and other renewable energies.

As the integration of wind power continues to grow, the concerns about its stability impacts outstand. Practically speaking, power system operations on one hand concern about the variability of wind power, whereas on the other hand regard the wind power as constant in stability analysis to reduce the computational cost. The theory-based method proposed in the paper takes the statistic properties of wind power into account, and meanwhile does not increase the computational burden. Specifically, the computational cost for the proposed method is the same as the model using constant wind speed, which has been widely applied \cite{Vieira:2015}-\cite{Loparo:2011}; the proposed deterministic model can also be easily incorporated in any dynamic model originally used in the time domain simulation. 
In addition, the analytical results in the paper provide an answer to the question that when the variability of wind power needs to be considered in long-term stability study, and when it does not. Those results will give a practical guidance for power system stability assessments if efficient numerical schemes can be developed to check the sufficient conditions presented in \textit{Theorem 2}. For other future works, an extension of the proposed method to the stochastic long-term stability model with discrete variables is expected. 
In addition, the proposed method can be further improved by applying the QSS model \cite{Cutsem:book}\cite{Cutsem:article2000}. As such, the QSS model is able to incorporate wind generation into the formulation, and thus provides a more efficient computational tool to conduct long-term stability analysis for power grids with wind power.
This extension of the QSS model may be a valuable advance in dynamic stability analysis, considering the wide industry applications of the QSS model and the continuous increasing of wind penetration. The analytical results proposed in this paper will be the foundation for those exciting subsequent works.

\appendices
\section{singular perturbation theory for SDE}\label{appendix1}
We consider the following stochastic system:
\begin{eqnarray}\label{sto_singular perburbed}
d{z}&=&f(z,x)dt+\bar{\sigma}Q(z,x)dW\qquad\quad z\in\mathbb{R}^{n_z}\\
d{x}&=&\frac{1}{\ee}g(z,x)dt+\frac{\sigma}{\sqrt{\ee}}G(z,x)dW\qquad x\in\mathbb{R}^{n_x}\nonumber
\end{eqnarray}
where $W$ denotes a $k$-dimensional standard Wiener process, and $Q\in \mathbb{R}^{n_z\times k}$, $G\in\mathbb{R}^{n_x\times k}$ are matrix-valued functions. We are interested in the case that $\bar{\sigma}$ does not dominate $\sigma$, i.e., $\bar{\sigma}=\rho \sigma$ where $\rho$ is uniformly bounded above in $\ee$.

\textit{Assumption 2.}

\textit{2.1} In some open connected set $D\subset \mathbb{R}^{n_z}\times\mathbb{R}^{n_x}$, $f\in C^2(D,\mathbb{R}^{n_z})$, $g\in C^2(D,\mathbb{R}^{n_x})$ and $Q\in C^1(D,\mathbb{R}^{n_z\times k})$, $G\in C^1(D,\mathbb{R}^{n_x\times k})$. $f$, $g$, $Q$, $G$ and all their partial derivatives up to order 2, respectively 1, are uniformly bounded in norm in $D$. 

\textit{2.2} The slow manifold $x=j(z)$ where $z\in D_0\subset \mathbb{R}^{n_z}$ of deterministic system (\ref{singular perturb}) is a stable component of constraint manifold $\Gamma$.

\textit{2.3} The diffusion matrix $G(z,j(z))G(z,j(z))^T$ is uniformly positive definite in $z\in D_0$.

Theorem 2.4 in \cite{Gentz:2003} which is reviewed as \textit{Theorem 3} as below states that under Assumption 2, the sample path of system (\ref{sto_singular perburbed}) are concentrated in an ellipsoidal layer surrounding the invariant manifold $j^\star(z,\ee)$ of system (\ref{singular perturb}):
\begin{equation}
M(h)=\{(z,x):\langle (x-j^\star(z,\ee)),J^\star(z,\ee)^{-1}(x-j^\star(z,\ee))\rangle< h^2 \}
\end{equation}
up to time $t$, with a probability roughly like $1-(t^2/\ee)e^{-h^2/2\sigma^2}$. In particular, $J^\star(z,\ee)$ describing the cross section of the ellipsoid is well defined under Assumption 2. Refer to Section 2.3 of \cite{Gentz:2003} or Chapter 5.1.1 of \cite{Gentz:2006} for more details.
%

The following theorem shows that the sample paths of the stochastic system (\ref{sto_singular perburbed}) are concentrated in a $\sigma$-neighborhood of invariant manifold $j^\star(z,
\ee)$ of the deterministic system (\ref{singular perturb}).



\textit{Theorem 3\cite{Gentz:2003}:} Under Assumption 2, for some fixed $\ee_0>0$, $h_0>0$, there exist $\delta_0>0$, a time $\tau_1$ of order $\ee|\mbox{log}h|$ such that whenever $\delta\leq\delta_0$, if an initial condition of stochastic system (\ref{sto_singular perburbed}) $(z(0),{x}(0))$ where $z(0)\in D_0$ satisfies $(z(0),{x}(0))\in M(\delta)$, then for $\tau\geq\tau_1$, the following relation
\begin{eqnarray}\label{upperbound}
&&\mathbb{P}\{\exists s\in[\tau_1,\tau]:(z(\tau),{x}(\tau))\notin M(h)\}\nonumber\\
&&\leq C_{n_{z},n_x}(\tau,\ee)e^{\frac{-h^2}{2\sigma^2}(1-O(h)-O(\ee))}
\end{eqnarray}
holds for all $\ee\leq\ee_0$, $h\leq h_0$, where coefficient $C_{n_{z},n_x}(\tau,\ee)=[C^{n_{z}}+h^{-n_x}](1+\frac{\tau}{\ee^2})$ is linear time-dependant.


A matching lower bound given in Theorem 2.4 of \cite{Gentz:2003} shows that the above bound is sharp in the sense that it captures the correct behaviour of the probability.


\textit{Remark 1}: It is shown in Theorem 5.1.17 in \cite{Gentz:2006} (Remark 2.7 and Lemma 3.4 in \cite{Gentz:2003}) that the stochastic singular perturbed system (\ref{sto_singular perburbed}) can be approximated 
by the reduced deterministic system:
\begin{equation}\label{det slow}
dz=f(z,j^\star(z,\ee))dt
\end{equation}
and the deviation between the solution of (\ref{det slow}) and that of (\ref{sto_singular perburbed}) is of order $\sigma\sqrt{\ee}$ up to Lyapunov time of system (\ref{det slow}). That means $z_D(t)$ provides $O(\sigma\sqrt{\ee})$ approximation for $z(t)$, i.e., $|z(t)-z_D(t)|=O(\sigma\sqrt{\ee})$.



\section{cross section of $N(h)$}\label{appendix2}
Following the deduction of $J^\star(z,\ee)$ in Section 2.3 of \cite{Gentz:2003} or Chapter 5.1.1 of \cite{Gentz:2006}, we can readily obtain the following results. The cross section $L_1^\star(z_c,z_d,\ee)$ of $N(h)$ is the solution of the following slow-fast system:
\begin{eqnarray}\label{transformedpower}
{z_c}^\prime&=&H_c(z,l_1^\star(z_c,z_d,\ee))\label{n(h)}\\
\ee{L_1}^\prime&=&A(z_c,z_d,\ee)L_1^\star+L_1^\star A(z_c,z_d,\ee)^T+\left[\begin{array}{cc} \kappa I_{n_{{x}}}&0\\0& I_{n_w}\end{array}\right]\nonumber
\end{eqnarray}
where
\begin{eqnarray}
A(z_c,z_d,\ee)&=&\partial_{\bar{x}} F(z_c,{l_1^\star}(z_c,z_d,\ee))\\
&&-\ee \partial_{z_c}{l_1^\star}(z_c,z_d,\ee)\partial_{\bar{x}} H_c(z_c,l_1^\star(z_c,z_d,\ee))\nonumber
\end{eqnarray}
and $\kappa>0$ is a sufficiently small positive parameter.


Note that the exact expression of $L_1^\star(z_c,z_d,\ee)$ is not of interest and thus does not need to be calculated, and we only need to know that $L_1^\star(z_c,z_d,\ee)$ is well defined in this formulation.


\section{proof of theorem 2}\label{appendix3}
\textit{Proof:} According to \textit{Theorem 2}, if the conditions that $h\gg\sigma$ and (i)-(ii) in \textit{Theorem 2} are satisfied, then after a time of order $\ee|\mbox{log}h|$, sample paths of the stochastic long-term stability model (\ref{sde power}) are concentrated in $N(\sigma)$ which is a $\sigma$-neighborhood of the invariant manifold $l_1^\star(z_c,z_d,\ee)$. 
Particularly, the slow dynamics of the stochastic long-term stability model (\ref{sde power}) are in the $O(\sigma\sqrt{\ee})$ neighborhood of the invariant manifold $l_1^\star(z_c,z_d,\ee)$ as explained in \textit{Remark 1}.
Hence, Eqn (\ref{corollary3_1}) and Eqn (\ref{corollary3_2}) hold only if trajectory of the deterministic long-term stability model moves along the invariant manifold $l_1^\star(z_c,z_d,\ee)$. On the other hand, from singular perturbation theory, we have that if the slow manifold $\bar{x}=l_1(z_c,z_d)$ is a stable component of the constraint manifold, then the invariant manifold $l_1^\star(z_c,z_d,\ee)=l_1(z_c,z_d)+O(\ee)$ exists for sufficiently small $\ee$. Moreover, if one more condition that the initial point of the deterministic long-term stability model lies inside the stability region of the initial short-term stability model is also satisfied,  
then there exists $\ee_0$ such that trajectory of the deterministic long-term stability model (\ref{det power}) moves along the invariant manifold $l_1^\star(z_c,z_d,\ee)$ for all $\ee\in(0,\ee_0)$,  $\tau\in[\tau_1,\tau_2]$\cite{Khalil:book}\cite{Wangxz:CAS}. Thus Eqn (\ref{corollary3_1}) and Eqn (\ref{corollary3_2}) hold for all $\ee\in(0,\ee_0)$, $\tau\in[\tau_1,\tau_2]$. This completes the proof.

\ifCLASSOPTIONcaptionsoff
  \newpage
\fi

\end{document}